\documentclass{jaa}
\def\lsim{\lower.5ex\hbox{$\; \buildrel < \over \sim \;$}}
\def\gsim{\lower.5ex\hbox{$\; \buildrel > \over \sim \;$}}
\def\be{\begin{equation}}
\def\bea{\begin{eqnarray}}
\def\eea{\end{eqnarray}}
\def\ee{\end{equation}}

\def\md{\dot {\cal M}}

\def\mbh{M_{\rm B}}
\def\bc{\begin{center}}
\def\ec{\end{center}}
\def\eg{{\it e.g.,}}
\def\etal{{\em et al.}}
\def\ie{{\em i.e.,}}

\def\rg{r_{\rm g}}

\def\tg{t_{\rm g}}
\def\xsh{x_{\rm sh}}
\def\Hsh{H_{\rm sh}}

\usepackage{color}
\usepackage{graphicx}

\begin{document}

\title{Radiatively driven general relativistic jets}


\author{Mukesh K. Vyas\textsuperscript{1} \and Indranil Chattopadhyay\textsuperscript{1}}
\affilOne{\textsuperscript{1}Aryabhatta Research Institute of Observational Sciences (ARIES), Manora Peak, Nainital-263002, India
\\}


\twocolumn[{

\maketitle

\corres{indra@aries.res.in}

\msinfo{}{}{}

\begin{abstract}
We use moment formalism of relativistic radiation hydrodynamics to obtain equations of motion of radial jets and solve them using polytropic equation of state of the relativistic gas. We consider curved space-time around black holes and obtain jets with moderately relativistic terminal speeds. In addition, the radiation field from
the accretion disc, is able to induce internal shocks in the jet close to the horizon. Under combined effect of thermal as well as radiative driving, terminal speeds up to 0.75 (units of light speed) are obtained.
\end{abstract}

\keywords{Radiation hydrodynamics, Hydrodynamics, Shocks, Black holes, Jets and outflows}
}]


\doinum{12.3456/s78910-011-012-3}
\artcitid{\#\#\#\#}
\volnum{123}
\year{2016}
\setcounter{page}{23}

\section{Introduction}
Jets are ubiquitous in astrophysical objects like active galactic
nuclei (AGN e.g., M87), young stellar 
objects (YSO e.g., HH 30, HH 34), X-ray binaries ({\eg} SS433, Cyg
X-3, GRS 1915+105, GRO 1655-40) etc. \\
This paper is devoted to dynamics of relativistic jets around black hole (hereafter BH) candidates like BH X-ray binaries. In such systems, jets can only emerge from accreting matter as BHs neither have hard surface nor they are capable of emission. This fact is supported by strong correlation observed between spectral state of the accretion disc and jet
(Fender \etal 2010; Gallo \etal 2003; Rushton \etal 2010).
Observations also shows that the jet generation region is very close, less than $100$
Schwarzschild radii ($r_{\rm s}$)
around the central object (Junor \etal. 1999; Doeleman \etal. 2012). 
This implies that entire accretion disc doesn't take part in jet generation. Following this, we assume that jets are launched within the accretion funnel close to the BH. Further, numerical simulations (Molteni \etal 1996; Das \etal 2014; Lee \etal 2016) and theoretical studies (Chattopadhyay \& Das 2007; Das \& Chattopadhyay
2008; Chattopadhyay \& Kumar 2016; Kumar \etal. 2013; Kumar \& Chattopadhyay 2017) showed that additional thermal gradient term in the accretion corona is able to give rise to bipolar outflows close to the BH.

After the very first theoretical model of accretion discs (Shakura \& Sunyaev 1973)
being Keplerian in nature, there have been numerous attempts to understand the interaction of radiation with jets (Icke 1980; Sikora \& Wilson 1981; Paczy\'nski \& Wiita 1980; Fukue 1996; Chattopadhyay \& Chakrabarti 2000a, Chattopadhyay \& Chakrabarti 2000b, Chattopadhyay \& Chakrabarti 2002, Chattopadhyay \etal 2004, Chattopadhyay 2005). Most of these attempts were made considering jets in particle regime while observations reveal their fluid nature. Ferrari \etal (1985) studied isothermal and non-radial fluid jets under Newtonian gravity having arbitrary radiation field in special relativistic regime. They obtained mildly relativistic jets and shocks induced by non radial nature of the cross section.
Isothermal assumption does not contain the effect of the thermal gradient term which is a significant accelerating agent and is very effective close to the BH. It is also the same region where one needs to consider the effects of general relativity as well. In this series Vyas \etal (2015) studied special relativistic jets under radiation field considering pseudo-Newtonian gravitational potential. They obtained relativistic jets but no multiple sonic points or shock transition was obtained. This paper extends the work of Ferrari \etal (1985) and Vyas \etal (2015) by considering fluid jets in curved space-time. Further, as Vyas \& Chattopadhyay (2017) showed that non radial jets, even without radiation field do create internal shocks while radial jets do not. Here we explore the possibility that radial jets can form shocks under the impact of sufficiently intense radiation field. \\

The equations of motion of radiation hydrodynamics were developed 
by many authors (Hsieh \& Spiegel 1976; Mihalas \& Mihalas 1984) and later their general relativistic version was obtained in further studies (Park \etal 2006, Takahashi 2007). 
In this paper we mainly follow the moment formalism to calculate the radiation field above accretion disc (Park \etal 2006).

In next section \ref{sec2}, we present assumptions, equations of motion and brief account of the
procedure to compute the radiation field. The methodology to obtain solutions is narrated in \ref{sec:method}.
Finally, we present results and draw conclusions in section \ref{results}
\begin {figure}[h]
\begin{center}
 \includegraphics[width=8.cm, trim=0 0 50 50,clip]{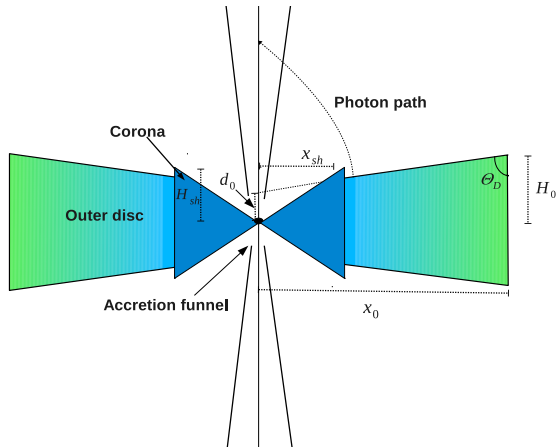}
\vskip -0.5cm
 \caption{Cartoon diagram of cross-sections of axis-symmetric
accretion disc and the associated jet in ($r,~\theta, \phi$ coordinates).
The outer limit of corona
$\xsh$, the intercept of outer disc on the jet axis ($d_0$), height of the corona $H_{\rm sh}$, the outer edge of the
disc $x_0$ are marked.}
\label{lab:fig1}
 \end{center}
\end{figure}

\section{Assumptions and governing equations}
\label{sec2}
\subsection{Assumptions}

We invoke general relativity
to take care of space-time curvature, which around a non-rotating BH is described by Schwarzschild metric:
\bea
ds^2=-g_{tt} c^2dt^2+
g_{rr}dr^2
+g_{\theta \theta}d{\theta}^2+g_{\phi \phi}d\phi^2  \nonumber \\
=-\left(1-\frac{2G\mbh}{c^2r} \right)c^2dt^2+
\left(1-\frac{2G\mbh}{c^2r}\right)^{-1}dr^2 \nonumber \\
+r^2d{\theta}^2+r^2\sin^2{\theta}d\phi^2
\label{metric.eq}
\eea

Here $r$, $\theta$ and $\phi$ are usual spherical 
coordinates,
$t$ is time, $g_{\mu \mu}$ are diagonal metric components, $\mbh$ is the mass of the central 
black hole and $G$ is the universal constant of gravitation. Hereafter, we have used geometric units (unless specified otherwise) with $G=M_B=c=1$ with the units of mass, length and time being $M_B$, $\rg=GM_B/c^2$ and $\tg=GM_B/c^3$ respectively for which, the event horizon is at $r_{\rm \small S}=2$. The jet is assumed to be in steady state ({\ie} $\partial/\partial t=0$) and as the relativistic jets are collimated, we consider on-axis ({\ie} $u^r=u^{\phi}=\partial/\partial r=0$) and axis-symmetric ($\partial/\partial \phi=0$) jet with small opening angle. Narrow jet allows us to further assume that at distance $r$, the physical variables of the jet remain same along its breadth. The jet is assumed to expand radially, perpendicular to the accretion plane. Further, a jet should have low angular momentum else it cannot remain collimated and following the effective angular momentum removal by radiation and magnetic fields, we assume jets to be non-rotating.
The cartoon diagram of disc jet system is shown in figure (\ref{lab:fig1}). The accretion disc has an outer disc and the inner torus like corona. The outer edge of corona and inner edge of outer disc is presented by
$\xsh$. The height of the corona is assumed to be $H_{\rm sh}=2.5 \xsh$.
Accretion disc works as a source of radiation emitting via synchrotron, bremsstrahlung and inverse Compton processes along with assumption that magnetic pressure in the disc is a fraction $\beta$ of the gas pressure. We take $\beta=0.5$ in this paper. To compute the radiation from the disc,
the density, velocity and the temperature distribution of the accretion disc has to be estimated.
We follow the methods of Vyas et. al. (2015) to obtain an analytical estimate of the flow variables in the accretion disc. We do not consider how the jets are being launched from the jet. The accretion disc plays an auxiliary role only. The plasma is assumed to be fully ionized and the interaction between radiation and matter is dominated by Thomson scattering. In scattering regime only momentum is transferred between radiation and matter and no energy transfer takes place. The relativistic effects on radiation field observed are also incorporated. The relativistic effects in the radiative transfer explicitly appears in the equations of motion while the effects of photon bending in radiation field are approximated taking the help of Beloborodov (2002); Bini \etal 2015.
Beloborodov's (2002) analysis approximated transformed radiation field due to curved space-time which is close to the exact values.
The transformation of  flat space-time relativistic specific intensities ($I_{j_f}$) into curved
space-time are given as  
\begin{equation}
I_j=I_{j_f}\left(1-\frac{2}{r_a}\right)^2
\label{Itrans.eq}
\end{equation}
Here $r_a$ is the radial coordinate of the source point on the accretion disc. 
The suffix $j{\rightarrow \rm {\small OD,~C}}$ signifies the
contribution from the outer disc and the corona, respectively. The square of redshift factor $(1-2/r_a)$ shows that curved space-time reduces the observed intensity.

Further, as photon moves in curved path, the transformed expressions of the direction cosine and solid angle
are given in terms of their flat space counterparts as (Beloborodov 2002),

\bea
l_j=l_{jf}\left(1-\frac{2}{r_a}\right)+\frac{2}{r_a} \nonumber \\
d \Omega_j = \left(1-\frac{2}{r_a}\right) d \Omega_{jf}
\label{l_do_trans.eq}
\eea
Using these transformation laws, the radiative moments namely radiation energy density ($R_0$), radiation flux ($R_1$), radiation pressure ($R_2$) and disc luminosities are calculated using similar procedure as in Vyas \etal (2015). Here we have excluded radiation contribution from Keplerian disc as its contribution in the
various components of
total radiative moments was found to be negligible.
The luminosity of the {\small OD} is obtained by integrating specific intensities over the disc surface and then using the luminosity ratio relation between {\small OD} and {\small C} (Vyas \etal 2015) for $M_B=10M_\odot$ we obtain luminosities of {\small C} or corona. The total luminosity ($\ell$) of the disc then is addition of both luminosities and is shown in units of Eddington luminosity in this paper. We treat $\ell$ as an input parameter. 

\subsection{Governing equations}

\subsubsection {Equation of state}
\label{sbsbsec2.1.1}
Equation of state (EoS) is a
closure relation between internal energy density ($e$), pressure ($p$) and mass
density ($\rho$) of the fluid.
In this study, we consider the jet fluid obeying polytropic EoS having fixed adiabatic index ($\Gamma=1.5$) 
given as,
\begin{equation}
e=\rho+\frac{p}{\Gamma-1}
\label{eos.eq}
\end{equation}

Expressions for adiabatic sound speed $a$ in relativistic regime and enthalpy $h$ are given by
\begin{equation}
a^2=\frac{\Gamma p}{e+p}=\frac{\Gamma \Theta}
{1+N \Gamma \Theta}; ~~ h=\frac{e+p}{\rho}=1+\Gamma N \Theta
\label{sound.eq}
\end{equation}
Here $N(=\frac{1}{\Gamma-1}=2)$ is polytropic index of the flow and non-dimensional temperature is defined as
$\Theta=p/\rho$.
\subsubsection {Dynamical equations of motion}
In relativistic notation the equations of motion of the any system are obtained when the four divergence of the energy-momentum tensor $T^{\alpha \beta}=T^{\alpha \beta}_{R}+T^{\alpha \beta}_{M}$ is set to zero. \ie 

\begin{equation}
T^{\alpha \beta}_{;\beta}=(T^{\alpha \beta}_{R}+T^{\alpha \beta}_{M})_{;\beta}=0
\label{four_d.eq}
\end{equation}
Here, $T^{\alpha \beta}_R$ and $T^{\alpha \beta}_M$ stand for jet matter and radiation field respectively and are given by (Mihalas \& Mihalas 1984)
\begin{equation}
T^{\alpha \beta}_M=(e+p)u^{\alpha}u^{\beta}+pg^{\alpha \beta};
~~T^{\alpha \beta}_R={\int}I_{\nu}l^{\alpha}l^{\beta}d{\nu}d{\Omega},
\end{equation}
The metric
tensor components are given by $g^{\alpha \beta}$, $u^{\alpha}$ are the components of four velocity,
$e$ and $p$ the fluid energy density, pressure in local co-moving frame.
Furthermore, $l^{\alpha}$s are the directional derivatives, $I_{\nu}$ is the specific intensity
of the radiation field with $\nu$ being the frequency of the radiation.
$\Omega$ is the solid angle subtended by a source point at the accretion disc surface on to the field point at the jet axis.

In absence of particle creation/destruction, conservation of four mass-flux is given by,
\begin{equation}
(\rho u^{\beta})_{; \beta}=0,
\label{eqnmot.eq}
\end{equation}
where, $\rho$ is the mass density of the fluid.
From above set of equations (eq. \ref{four_d.eq}), the momentum balance equation in the $i^{th}$ direction is obtained using projection tensor,
$(g^{i}_{\alpha}+u^iu_{\alpha})$. 
{\ie}
\begin{equation}
(g^{i}_{\alpha}+u^iu_\alpha)T^{\alpha \beta}_{M_{;\beta}}=-(g^{i}_{\alpha}+u{^iu_\alpha})
T^{\alpha \beta}_{R_{;\beta}}
\label{genmomb.eq}
\end{equation}
For an on axis jet in steady state it becomes (Park \etal 2006)
\begin{equation}
u^r\frac{du^r}{dr}+\frac{1}{r^2}=-\left(1-\frac{2}{r}+u^ru^r\right)
\frac{1}{e+p}\frac{dp}{dr}+\frac{\rho_e{\sigma}_{T}
}{m_p(e+p)}{\Im}^r,
\label{eu1con.eq}
\end{equation}
Here $\rho_e$ is lepton mass density, $m_p$ is the mass of the proton
and ${\Im}^r$ is the net radiative contribution and is given by;

\be
{\Im}^r=\sqrt{g^{rr}}\gamma^3\left[(1+v^2){R_1}-v
\left(g^{rr} R_0+\frac{R_2}{g^{rr}}\right)\right]
\ee

Here we define three velocity $v$ of the jet as 
$$
v^2=-u_iu^i/u_tu^t=-u_ru^r/u_tu^t \Longrightarrow u^r={\gamma}v\sqrt{g^{rr}}
$$
and $\gamma^2=-u_tu^t$ is the Lorentz factor. $R_0, R_1$ and $R_2$ are zeroth, first and second moments of specific intensity. 
Similarly, the energy conservation equation is obtained by taking 
\begin{equation}
u_{\alpha}T^{\alpha \beta}_{M_{;\beta}}=-u_{\alpha}T^{\alpha \beta}_{R_{;\beta}}
\label{genfstlaw.eq}
\end{equation}

In the scattering regime, it becomes 
\begin{equation}
\frac{de}{dr}-\frac{e+p} {\rho}\frac{d\rho}{dr}=0,
\label{en1con.eq}
\end{equation}
Absence of emission/absorption makes the right side of equation (\ref{en1con.eq}) zero. It is a consequence of scattering regime assumption and shows that the system is isentropic.
From continuity equation (eq. \ref{mdotout.eq}) the mass outflow rate is given as 
\begin {equation}
\dot {M}_{out}=\Omega \rho u^r r^2;~~ \Omega \mbox{=geometric constant}
\label{mdotout.eq}
\end {equation}
The differential form of the outflow rate equation is, 
\begin{equation}
\frac{1}{{\rho}}\frac{d{\rho}}{dr}=-\frac{2}{r}
-\frac{1}{u^r}\frac{du^r}{dr}.
\label{con1con.eq}
\end{equation}

Equation (\ref{en1con.eq}) can be integrated with help of equation (\ref{eos.eq}) to obtain
isentropic relation between $p$ and $\rho$,
$$
p=k\rho^{\Gamma}
$$
where, $k$ is entropy constant of the flow.
This equations enables us to replace $\rho$ from equation (\ref{mdotout.eq}), and we obtain the expression for entropy-outflow
rate as,
\begin{equation}
\md=\Theta^{N}u^rr^2
\label{entacc.eq}
\end{equation}
$\md$ remains constant along the streamline of the jet, except at the shock.

Integrating equation (\ref{eu1con.eq}), we obtain generalized, relativistic
Bernoulli parameter for the radiatively driven jet,
\be 
E=-h u_t{\rm exp}\left(\int dr \frac{\sigma_T(1-Na^2)\Im^r}{m_p \gamma^2(1-2/r)}\right)
\label{energy.eq}
\ee
Momentum balance equation (eq. \ref{eu1con.eq}), with the help of equation (\ref{con1con.eq}), 
is simplified to
\bea
\gamma^2vg^{rr}r^2\left(1-\frac{a^2}{v^2}\right)\frac{dv}{dr}=a^2\left(2r-3\right)-1 \nonumber \\
+\frac{\Im^r r^2(1-Na^2)}{m_p \gamma^2}
\label{dvdr.eq}
\eea
Using energy conservation equation (\ref{en1con.eq}) along 
with the EoS (eq. \ref{eos.eq}), the expression of temperature gradient along $r$ is obtained to be 
\begin{equation}
\frac{d{\Theta}}{dr}=-\frac{{\Theta}}{N}\left[ \frac{{\gamma}
^2}{v}\left(\frac{dv}{dr}\right)+\frac{2r-3}{r(r-2)}
\right]
\label{dthdr.eq}
\end{equation}

Equations (\ref{four_d.eq}) and (\ref{eqnmot.eq}), the two equations of motion reduces to
two differential equations (\ref{dvdr.eq}) and (\ref{dthdr.eq}), which describes the distribution
of two flow variables $v$ and $\Theta$. 
\section{Methods of analysis}
\label{sec:method}
The solution for radiatively driven jet can be obtained if equations (\ref{dvdr.eq}) and (\ref{dthdr.eq}) are solved. Since jets are launched from accretion disc, close to the central object, so the injection speed will
be small, while the temperature will be high. So at the base, jets should be subsonic.
Far away from the base, jets are observed to be moving with relativistic speed and therefore supersonic. Hence such flows are transonic in nature. The distance ($r=r_c$) at which the bulk speed ($v=v_c$) crosses the local sound speed ($a=a_c$), is called the sonic point.
Equation (\ref{dvdr.eq}) shows that the sonic point is also critical point since
at $r=r_c$, $dv/dr\rightarrow 0/0$. This property gives the sonic or critical point conditions,
\be 
v_c=a_c;
\label{sonic1.eq}
\ee
and 
\be 
a_c^2\left(2r_c-3\right)-1+\frac{\Im^r_c r_c^2(1-N_ca_c^2)}{m_p \gamma_c^2}=0
\label{sonic2.eq}
\ee
Suffix $c$ denotes that the values are to be obtained at the sonic point ($r=r_c$).  For a given $r_c$,
we solve equation (\ref{sonic2.eq}) to find $a_c$ and then $\Theta_c$ (equation \ref{sound.eq}).
We can also
compute $\md_c$, $E_c$ at $r_c$ (using equations \ref{entacc.eq} and \ref{energy.eq}).
Since $E_c=E$ for a particular solution, therefore, for a given $E$, $r_c$ is determined and vice versa.
In other words, sonic point is a mathematical boundary. So we first obtain all the variables at $r_c$ and then calculate $|dv/dr|_c$ by using the L'Hospital's rule in equation (\ref{dvdr.eq}) at $r=r_c$. This leads to a quadratic equation for $|dv/dr|_c$, which can admit two complex roots having `spiral' type sonic points, or two real roots but with opposite signs (called $X$ or `saddle' type sonic points), or real roots with same sign (known as nodal type sonic point). For a given boundary values at the base of the jet ($r=r_b=3$) the transonic solutions will pass through sonic points determined by $E$ and $\md$ of the flow giving the values at the outer boundary $r_{\infty}$ (defined by $r=r_\infty=10^5$). 
We integrate equations (\ref{dvdr.eq} and \ref{dthdr.eq}) simultaneously inward and outward from the $r_c$ using $4^{th}$ order Runge–Kutta method. 

\subsection {Shock conditions}

The existence of multiple 
sonic points in the flow opens up the possibility of 
formation of shocks in the flow. At the shock,
the flow is discontinuous in density, pressure and velocity.
The relativistic Rankine-Hugoniot conditions relate the flow quantities across the
shock jump (Chattopadhyay and Chakrabarti 2011)
\begin{equation}
  [{\rho}u^r]=0,
  \label{sk1.eq}
\end{equation}
\begin{equation}
   [\dot{E}]=0
   \label{sk2.eq}
\end{equation}
and
\begin{equation}
[T^{rr}]=[(e+p)u^ru^r+pg^{rr}]=0
\label{sk3.eq}
\end{equation}
The square brackets denote the difference 
of quantities across the shock, i.e. 
$[Q]=Q_2-Q_1$ 
with $Q_2$ and  $Q_1$ being 
the quantities after and before the shock respectively.

Equation (\ref{sk2.eq}) states that the energy flux remains 
conserved across the shock. Dividing 
(\ref{sk3.eq}) by (\ref{sk1.eq}) and a little 
algebra leads to
\be
(1+\Gamma N \Theta)u^r+\Theta g^{rr}=0
\label{sk5.eq}
\ee
We check for shock conditions (equations \ref{sk2.eq}, \ref{sk5.eq})
as we solve the equations of motion
of the jet.
\section{Results and discussion}
\label{results}
\subsection{Nature of radiation field}
\begin {figure}[h]
\begin{center}
 \includegraphics[width=8.cm, trim=0 0 200 200,clip]{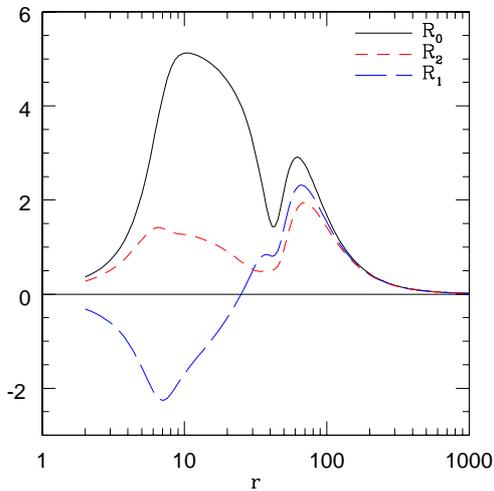}
\vskip -0.5cm
 \caption{Radiative moments $R_0, R_1$ and $R_2$ from accretion disc corresponding to accretion rate ($\dot{m}$) to be $9.145$ or disc luminosity ($\ell$) to be $0.5$. Both $\dot{m}$ and $\ell$ are in Eddington units.}
\label{lab:fig2}
 \end{center}
\end{figure}

In fig. (\ref{lab:fig2}) we show radiative moments $R_0$ (solid, black), $R_1$ (long-dashed, blue) and $R_2$
(red, dashed) as functions of $r$ calculated at the jet axis for $\dot{m}=9.145$ which corresponds to
$\xsh=14.67$ and $\ell=0.5$. The first peak ($\lsim 10$) in the moments, is due to the radiation
from the corona, and the second peak ($55$) is due to that from the outer disc.
Due to the shadow effect from the post shock disc, all moments from the outer disc are zero for $r<30$.
Since the corona is geometrically thick, the radiative flux $R_1$ is negative in the funnel like region.
The magnitude of the moments rise as the jet sees more of the disc as it propagates upward and they decay after reaching a peak value. The moments follow an inverse square law at large distances. The negative flux 
in the funnel pushes the jet material downward and works against the motion, to the extent that, it may drive
shock in jets.

\subsection{Nature of sonic points and behaviour of flow variables}
\begin {figure}[h]
\begin{center}
 \includegraphics[width=9.cm, trim=0 0 90 80,clip]{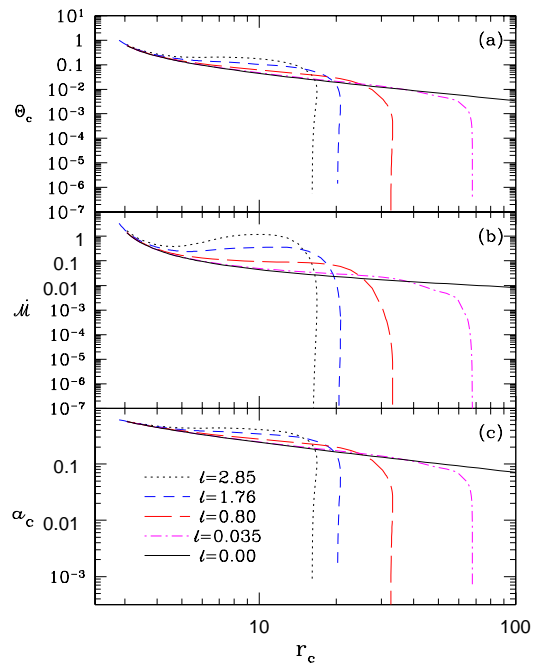}
\vskip 0.cm
 \caption{Variation of $\Theta_c$ (a), $\md_c$ (b) and $a_c$ (c) with $r_c$ for various values of $\ell$.}
\label{lab:fig3}
 \end{center}
\end{figure}

Using the procedure explained in section (\ref{sec:method}), we provide sonic point $r_c$ and calculate physical variables there. In Fig. (\ref{lab:fig3}a, b,and c) we show variation of $\Theta_c$, $\md_c$ and $a_c$ with
$r_c$, respectively. Various curves are plotted for different luminosities as $\ell=2.85$ (dotted, black),
$1.76$ (dashed, blue), $0.80$ (long-dashed, red), $0.035$ (dashed-dotted, magenta) and these all are compared with the thermal flow $\ell=0.00$ (solid, black). Physically different sonic points mean different choices of boundary conditions that give different transonic solutions, similarly, choice of an $r_c$ implies a solution with a unique choice of
$E$ and $\md$. For all possible values of $r_c$, thermal jets harbour real roots of $a_c$ (or, corresponding $\Theta_c$). While radiation field limits the region where $a_c$ can have real values.
For $r_c>r_{c_{\rm max}}$, one obtains complex values of $a_c$ (equation \ref{sonic2.eq}). It is also
found that $r_{c_{\rm max}}$ always lies inside the corona funnel (\ie $r_{c_{\rm max}}<H_{\rm sh}$). Physically, the critical points where $a_c$ is found imaginary, correspond to solutions where fluid approximation breaks down or physical temperatures are not defined. 
\begin {figure}[h]
\begin{center}
 \includegraphics[width=9.cm, trim=0 0 0 220, clip]{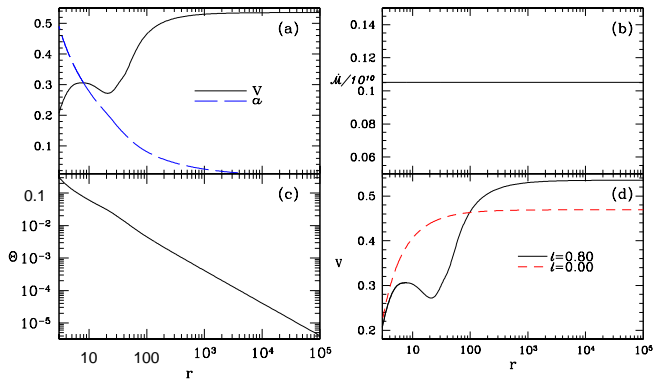}
\vskip -0.5cm
 \caption{(a) Variation of three velocity $v$ (solid black) and $a$ (long-dashed blue), (b) $\md$, 
and (c) $\Theta$ with $r$ for $\ell=0.80$. (d)
Comparison of $v$ for $\ell=0.8$ (solid black) with $v$ for $\ell=0$ (red dashed).
For all the plots $E=1.43$.}
\label{lab:fig4}
 \end{center}
\end{figure}

In Fig. (\ref{lab:fig4}) we show a typical nature of flow variables along $r$ for $\ell=0.8$ and $E=1.43$. In Fig. (\ref{lab:fig4}a), we show variation of three velocity $v$ (solid, black) and $a$ (long-dashed, blue). The effect of negative flux is clearly seen as $v$ decreases inside the funnel and then it accelerates above it. In Fig. (\ref{lab:fig4}b), we plot the entropy outflow rate $\md$ which remains constant since scattering
is an isentropic process. Figure (\ref{lab:fig4}c), shows sharp decline of temperature due to adiabatic expansion. In Fig. (\ref{lab:fig4}d), we compare $v$ of radiatively driven jet (solid, black), and thermally driven solution (dashed, red), both having the same $E$. We see that radiative acceleration dominates over radiative drag and terminal speed of the jet is higher in presence of radiation field.

\begin {figure}[h]
\begin{center}
 \includegraphics[width=9.5cm, trim=10 0 00 200,clip]{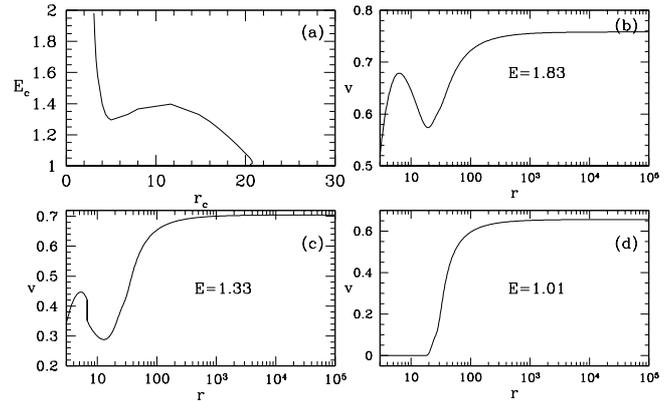}
\vskip -0.5cm
 \caption{(a) Variation of $E_c$ with $r_c$; Variation of $v$ with $r$ for (b) $E=1.83$, (c) $E=1.33$ and
 (d) $E=1.01$. For all the curves, $\ell=1.76$. }
\label{lab:fig5}
 \end{center}
\end{figure}
Now in figure (\ref{lab:fig5}a-d) we investigate the behaviour of jet speed with different boundary conditions, \ie different choices of $E$. We choose $\ell=1.76$ and plot $E_c$ with $r_c$ in Fig. (\ref{lab:fig5}a). For very high value of $E=E_c=1.83$ the flow is hot, and radiation is in-effective (Fig. \ref{lab:fig5}b). The jet accelerates due to the thermal gradient term and becomes transonic at $r_c=3.2$. As the jet expands, the temperature decreases and radiation becomes effective. The combined effect of negative flux in the
funnel and radiation drag term decelerates the speed. Above the funnel radiation flux become positive and starts
to accelerate the jet and it achieves terminal speed of $v_{\rm \small T}=0.76$.
If one chooses lower values of $E=1.33$ (Fig. \ref{lab:fig5}c), the jet passes through inner sonic point at $r=4.4$. Because the energy is low, radiation is more effective. Radiation flux opposes the
outflowing jet inside the funnel more vigorously and causes a shock transition --- a discontinuous transition from supersonic branch to subsonic branch at $r=6.78$ and then after coming out of the accretion funnel, it again accelerates under radiation push and becoming transonic forming an outer sonic point at $r=14.77$. The terminal speed achieved for this case is $\sim 0.7$. Here the jet crosses two sonic points with $\md$ to be higher for outer sonic point ($\md=0.291$) than the inner one ($\md=0.287$). Vyas \& Chattopadhyay (2017) showed that conical jets without radiation do not form shock, but here we see that radiation field is able to induce shocks in radially outflowing jets.  For even lower energy $E=1.01$ (Fig. \ref{lab:fig5}d),
the radiation is even more effective, and the jet speed is drastically reduced within the funnel. 
However, above the funnel it is accelerated very efficiently, becomes transonic through a single sonic point
and achieves terminal speed of about $v_{\rm \small T}\sim 0.65$.
 
\begin {figure}[h]
\begin{center}
 \includegraphics[width=9.cm, trim=0 0 00 250,clip]{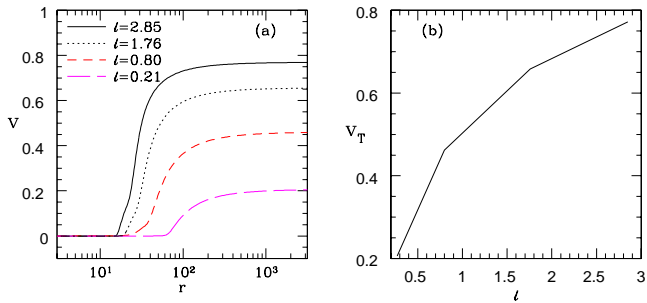}
\vskip -0.5cm
 \caption{ (a) Variation of $v$ with $r$ for various luminosities ranging from $\ell=0.21$ to 2.85 . (b) $v_T$ as a function of $\ell$.}
\label{lab:fig6}
 \end{center}
\end{figure}
We now choose low energy jets, whose base speeds are very low (similar to Fig. \ref{lab:fig5}d).
Since these jets have very low base speeds and base temperatures, we use them to compare jet speeds
acted on by various disc luminosities.  In Fig. (\ref{lab:fig6}a), we compare $v$ with $r$  for $\ell=2.85$ (solid, black), $\ell=1.76$ (dotted, black), $\ell=0.80$ (red, dashed) and $\ell=0.21$ (long-dashed, magenta).
We observe that higher radiation accelerates the jets up to greater speeds.
In Fig. (\ref{lab:fig6}b), we plot the terminal speeds ($v_{\rm \small T}$) as a function of $\ell$.
The base speed of these jets are very low.
For super Eddington luminosities, like $\ell=2.85$ the jet achieves terminal speeds to be around $0.75$. 
\subsubsection{Effect of corona geometry, magnetic pressure in disc ($\beta$) and $\Gamma$}
\begin {figure}[h]
\begin{center}
 \includegraphics[width=9.cm, trim=0 0 00 250,clip]{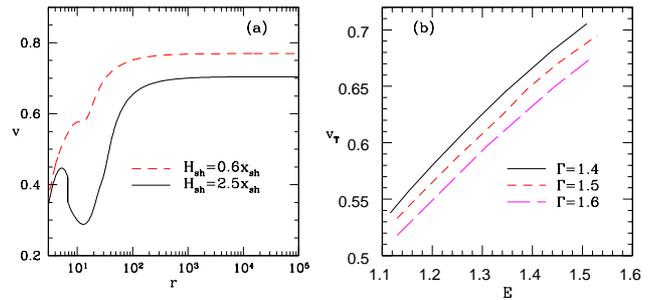}
\vskip -0.5cm
 \caption{ (a) Variation of $v$ with $r$ for various disc height ratios, $\Hsh=2.5\xsh$ (solid black) and $\Hsh=0.6\xsh$ (red dashed) for $\ell=1.76$ (b) $v_T$ as a function of $E$ for varying $\Gamma$ keeping $\ell=0.80$}
\label{lab:fig7}
 \end{center}
\end{figure}

In this paper we have considered thick discs with corona height being $2.5$ times it's width. One may wonder how the results would behave if discs are less thicker. To compare the effects of different geometries, we choose $\Hsh=0.6\xsh$ as in Vyas \etal (2015) and generate velocity profiles of the jet for $\ell=1.76$ and $E=1.33$ (same parameters as in Fig. \ref{lab:fig5}c). The profiles are plotted in Fig (\ref{lab:fig7}a) for thicker ($\Hsh=2.5\xsh$, solid, black) and thinner corona ($\Hsh=0.6\xsh$, dashed, red). It is clear that radiation from geometrically thick corona is more capable to produce shock, as the jet faces negative flux after being launched. For thinner corona, the radiation resistance is relatively less and the jet is unable to form shock inside the funnel.
Further, lesser resistance inside the funnel of thinner disc, makes radiative acceleration more effective and as a result the terminal speed is greater.

Choice of the value of $\Gamma$ is a tricky issue. This is because the base of the jet
is hot and $\Gamma$ should be closer to, but not exactly $4/3$ (Chattopadhyay \& Ryu, 2009).
And it should be lower than $5/3$, therefore, we took the median value
of $1.5$ in the previous sections. If one considers different values of $\Gamma$, then the behaviour
of the jet changes because different choices of $Gamma$ alter  the
net heat content of the flow..
In Figure (\ref{lab:fig7}b) we plot $v_T$ as a function of $E$ for $\ell=0.80$ and $\Gamma$($=1.4$, solid black), $\Gamma$($=1.5$, red dashed) and $\Gamma$($=1.6$, long dashed magenta). 
Smaller value of $\Gamma$ results in higher thermal driving and produces faster jets. 

\begin {figure}[h]
\begin{center}
 \includegraphics[width=8.5cm, trim=0 0 00 200,clip]{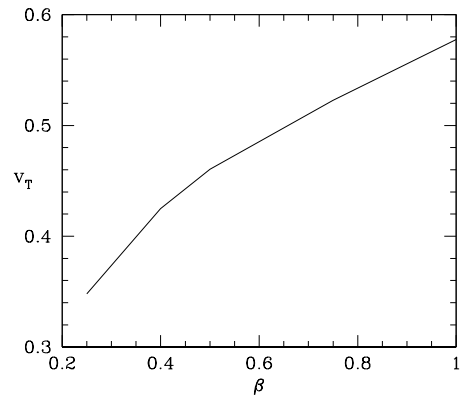}
\vskip -0.5cm
 \caption{$v_T$ is plotted as a function of $\beta$, for $\dot{m}=10$ and $\Hsh=2.5\xsh$.}
\label{lab:fig8}
 \end{center}
\end{figure}

In this study $\beta$ parameter is introduced to compute the synchrotron cooling from stochastic magnetic field.
Therefore in steady state it is most likely that $\beta < 1$ hence steady disc will not form.\
We took $\beta=0.5$ as an adhoc value. Increasing $\beta$ would increase synchrotron radiation, but would not
increase bremsstrahlung because $\dot{m}$ is not being changed. Moreover, the number of hot electrons which
inverse-Comptonize soft photons also do not change much, so although increasing $\beta$ amounts to increasing
$\ell$, but the distribution of $\ell$ is different and therefore, the response of $v_T$ to $\beta$
is different than $\dot{m}$ or $\ell$,
as was shown in Vyas et. al. (2015). 
In Fig. \ref{lab:fig8} shows the variation of $v_T$ as a function of $\beta$ for a given $\dot{m}$ and
$\Hsh=2.5\xsh$.
 
\subsection{Final Remarks}
This paper is development and expansion of earlier papers on radiatively driven fluid jets (Ferrari et. al. 1985;
Chattopadhyay \& Chakrabarti 2002, Vyas et. al. 2015). While the previous papers are either in Newtonian
or special relativistic regime, the present effort is in general relativistic regime. Only Ferrari et. al. (1985)
showed the presence of radiatively driven shocks in jets, but the terminal speeds were pitiable. Most likely,
the low terminal speeds are a result of their preference of isothermal approximation. Even though the details
of accretion disc physics have been ignored, but still broad geometric features like a thicker
corona and outer flatter disc, produced a significantly different radiative moment distribution, which resulted
in a set of very rich classes of jet solutions.


\end{document}